\documentstyle[preprint,aps,prb]{revtex}
\begin{document}
\draft

\title{Gutzwiller Approximation in Degenerate Hubbard Models}
\author{Jian Ping Lu}
\address{
Department of Physics and Astronomy \\
University of North Carolina at Chapel Hill \\
Chapel Hill, North Carolina 27599  \\ 
jpl@physics.unc.edu}

\date{\today}
\maketitle

\begin{abstract}
Degenerate Hubbard models
are studied using the Generalized-Gutzwiller-Approximation.
It is found that the metal-insulator transition occurs at
a finite correlation $U_c$ when the average
number of electrons per lattice site is an integer.
The critical $U_c$ depends
sensitively on both the band degeneracy $N$ and the filling $x$.
A derivation is presented for the
general expression of $U_c(x,N)$ which
reproduces all previously known Gutzwiller solutions,
including that of the Boson Hubbard model.
Effects of the lattice structure on the metal-insulator transition and
the effective mass are discussed.
\end{abstract}
\bigskip
\bigskip

The simple Hubbard model, where each lattice site is occupied
by at most two electrons, has been extensively
studied in recent years as a model of strongly correlated
electron systems.\cite{hubbard,montorsi}
But in many systems, such as molecular
solids $C_{60}$ and $A_xC_{60}$,\cite{a3c60} the 
conduction band is degenerate which allows the occupation of more
than two electrons per lattice site.
For these systems, studying the degenerate Hubbard model on a lattice
is the first step toward the understanding of electron-electronic correlations.
Yet, there has been little progress
since the pioneering works of Chao and Gutzwiller in early 1970s.\cite{chao}

Recently, we succeeded in obtaining the Gutzwiller-like analytical
solutions to degenerate Hubbard models using a 
scheme similar to the
original Gutzwiller-Approximation.\cite{gutz}
A brief report of our results has been published.\cite{dhub1}
The solution was used successfully to interpret the
unusual metal-insulator transitions observed
in fullerides $A_xC_{60}$.\cite{dhub1,mota}
In this paper we present details of our derivation and the approximation.
In addition, effects of the crystal lattice structure on
the metal-insulator transition and the effective mass
are also investigated.

We consider the $N$-fold degenerate Hubbard model defined on a lattice
\begin{equation}
H= \sum_{<ij>=n.n., \alpha} t c_{j,\alpha}^{+} c_{i,\alpha}
+ \frac{U}{2} \sum_{i,\alpha \ne \beta} n_{i,\alpha}n_{i,\beta} \, ,
\end{equation}
where $<ij>$ are nearest neighbor lattice sites, $\alpha=(r,\sigma)$ includes 
both the spin ($\sigma=\uparrow,\downarrow$) and the orbital 
($r=1,2,\cdots,N$) indices, and
$n_{i,\alpha}=c_{i,\alpha}^{+}c_{i,\alpha}$ is
the number operator. 
Both the Hubbard $U$ and the nearest neighbor
hopping $t$ are assumed to be independent of lattice sites, electron spins,
and molecular (atomic) orbitals. 
Let $L$ be the total number of lattice sites,
$x$ the average number of particles per site.
For simplicity, only total symmetric states are
considered. Thus, the number of electrons per orbital per spin is
$m=xL/2N$. For an integer $x$ one expects an insulating ground state
for $U>>t$.
In such an insulating state there are 
$x$ localized electrons at each lattice site.
Hopping between sites are quenched due to the large
correlation energy cost.
On the other hand, for $U=0$ the system is metallic as
electrons hop between
lattice sites to lower the kinetic energy.
One expects that there exists a critical $U_c$ across which
the system undergoes a metal-insulator transition.
In general, the insulating ground state can be magnetically ordered.
We will consider only non-magnetic states.
Though the qualitative results described below may
be valid for magnetically ordered states, it is clear that further
works are required to include the magnetic ordering.

If an electron hops from a site with $x$ particles to
a neighboring site which also has $x$ particles,
then the correlation energy cost,
$U(C_{x-1}^{2}+C_{x+1}^{2}-2C_{x}^{2})=U$, is independent
of the band filling $x$ and the degeneracy $N$. 
The gain in kinetic energy is of the order
of band width.
Thus, in a simple mean field argument 
one might expect that $U_c$ to be independent of
the band filling $x$ and the degeneracy $N$.
We will show that this is not true.
In fact the solution we obtained in a Gutzwiller-type approximation
predicts that $U_c$ depends sensitively on both $x$ and $N$ . 

Metal-insulator transition in non-degenerate
Hubbard models have been extensively studied in recent years.\cite{montorsi}
At the half-filling (one electron per site) it is known:
(1) In 1D,  the Bethe Ansatz exact solution\cite{liewu} shows that the ground state
is insulating for $U>0$.
(2) In 2D, the quantum Monte Carlo calculations
suggest an insulating ground state for a moderate $U$.\cite{scalapino}
In addition, for an extended Hubbard model
(the so called Kievelson-Schriffer-Su-Heeger model)
it has been shown exactly that the Mott-Hubbard transition exists
at a non-zero finite $U_c$,\cite{montorsi},
and the $U_c$ agrees with that obtained by the Gutzwiller-Approximation.
(3) In 3D or higher it is generally believed that a finite $U_c$
exists.
(4) In the infinite dimensions the Gutzwiller-Approximation is exact.
From these results we conclude that the Gutzwiller-Approximation
provides a reasonable estimate of the Mott-Hubbard
transition in 3D degenerate Hubbard models.

In the simple Hubbard model each lattice site are either
empty, singly-occupied, or doubly-occupied.
Within the Gutzwiller approach, one introduces a wave
function which is a linear combination of states with
different number of doubly-occupied sites with
weighting factors to be determined variationally.
To make analytical calculations possible, Gutzwiller
made two approximations:\cite{gutz}
{\it 1) In the thermodynamic limits, the expectation value of an operator
in the variational function is approximated by the largest
term contributing to it}. This approximation is justified by
the fact that the weighting factor, which includes
various combinatorial factors, is a sharply
peaked function.
{\it 2) In evaluating the expectation the spatial correlations of doubly-occupied 
sites are neglected}. 
This is equivalent to the mean field approximation.
For the $N$-fold degenerate Hubbard model, the problem
is more complex as each site can be occupied by $0$
to $2N$ number of electrons.
Without further approximation
the combinatorial counting becomes intractable.
However, close to the metal-insulator
transition one expects that large fluctuations from 
the average occupancy $x$ is small as
they cost large correlation energy. This leads us to
make the additional approximation:
{\it 3) Each lattice site is restricted to three possible states:
 ``empty'' (E) --- with $x-1$ electrons,
``singly-occupied'' (S) --- with $x$ electrons,
or ``doubly-occupied'' (D) --- with $x+1$ electrons.}
This approximation makes it possible to do variational
calculations following similar steps of the
original Gutzwiller-Approximation. 
We will refer to conditions 1), 2), 3) as the 
Generalized-Gutzwiller-Approximation (GGA) for
the degenerate Hubbard model.

Let $\Gamma,\Omega,\Lambda$ be the total number of sites for
types (D), (S), (E) respectively.
Approximation 3) and the conservation of particles require that
$L=\Gamma+\Omega+\Lambda$, $\Lambda=\Gamma$.
For a fixed $\Gamma$ , there are many possible ways
of participating $L$ lattice sites.
A given (D) site is distinguished by
identifying flavors $p_1,p_2,\ldots,p_{x+1} 
\in (1,2,3,\ldots,2N)$ of all $x+1$ particles on the site.
The numbers of lattice sites occupied by the
same configuration in general depends on the configuration.
For clarity, we will use the short-hand notation 
$\gamma_p$ $(p=1,2,3,\ldots,C_{2N}^{x+1})$
for this number.
There are $C_{2N}^{x+1}=\frac{(2N)!}{(x+1)!(2N-x-1)!}$ 
distinct possible configurations for each (D) sites.
Similarly we define $\omega_r$ and $\lambda_q$ for 
sites of type (S) and (E). Respectively, there are $C_{2N}^{x}$ and
$C_{2N}^{x-1}$ possible configurations for
these sites. One has
\begin{equation}
\sum_{p=1}^{C_{2N}^{x+1}} \gamma_p=\Gamma, \;\;
\sum_{r=1}^{C_{2N}^{x}} \omega_r=\Omega, \;\;
\sum_{q=1}^{C_{2N}^{x-1}} \lambda_q=\Lambda \, .
\end{equation}
Following Gutzwiller\cite{gutz}, one introduces
the weighting factor $0<\eta<1$ for each
(D) site, the GGA variational wavefunction $|\phi_{GGA}>$ can be written
as a linear combination of states with different
$\{\lambda_p\},\{\omega_r\},\{\lambda_q\}$,
\begin{equation}
|\psi_{GGA}> = \sum_{ \{\lambda_p\},\{\omega_r\},\{\lambda_q\} }
(\eta)^\Gamma A(\{\lambda_p\},\{\omega_r\},\{\lambda_q\}) 
|\phi_{\{\lambda_p\},\{\omega_r\},\{\lambda_q\}}> \, .
\end{equation}
The coefficient $A$ is proportional 
to the total number of possible ways of participating the
lattice given $\{\lambda_p\},\{\omega_r\},\{\lambda_q\}$,
\begin{eqnarray}
{\left| A(\{\gamma_p\},\{\omega_r\},\{\lambda_q\} )\right|}^2
&=&const \times
\prod_{p=1}^{C_{2N}^{x+1}} C_{L-\sum_{i<p}\gamma_i}^{\gamma_p} \times
\prod_{r=1}^{C_{2N}^{x}} C_{L-\Gamma-\sum_{i<r}\omega_i}^{\omega_r} \times
\prod_{q=1}^{C_{2N}^{x-1}-1}
C_{L-\Gamma-\Omega-\sum_{i<q}\lambda_i}^{\lambda_q}  \nonumber \\
&=&const \frac{L!}{ (\prod_{p=1}^{C_{2N}^{x+1}} \gamma_p ! ) \times
(\prod_{r=1}^{C_{2N}^{x}} \omega_r ! ) \times
(\prod_{q=1}^{C_{2N}^{x-1}} \lambda_q !) } \, ,
\end{eqnarray}
where, the $const$ is determined by the normalization condition.
Given the general wavefunction Eq.(3), the expectation value
of an operator is difficult to calculate.
Within the Gutzwiller-Approximation one assumes
that in the thermodynamical limit only the
largest term contributes significantly to the expectation.
For such an optimal term, the symmetry of the problem dictates that
\begin{eqnarray}
\gamma_p&=&\gamma=\Gamma/{C_{2N}^{x+1}} \, , \nonumber \\
\lambda_q&=&\lambda=\Lambda/{C_{2N}^{x-1}}=\Gamma /{C_{2N}^{x-1}} \, ,\\
\omega_r&=&\omega=\Omega/{C_{2N}^{x}}=(L-2\Gamma)/{C_{2N}^{x}} \, .\nonumber
\end{eqnarray}
The relationship between $\eta$ and $\Gamma$ is determined 
by the largest term in \mbox{$<\psi_{GGA}|\psi_{GGA}>$}. This leads to
\begin{eqnarray}
{\eta}^{2(\Gamma+1)} |A(\gamma+1,&& \gamma,\gamma,\cdots,\gamma;
\omega-1,\omega-1,\omega,\cdots,\omega; \lambda+1,\lambda,\lambda,\cdots,\lambda)|^{2} \nonumber  \\
&&={\eta}^{2\Gamma} |A(\gamma,\gamma,\gamma,\cdots,\gamma;
\omega,\omega,\omega,\cdots,\omega; \lambda,\lambda,\lambda,\cdots,\lambda)|^{2} \, .
\end{eqnarray}
Using Eq.(2), (4) and (5) one obtains
\begin{equation}
\eta=\frac{\sqrt{\gamma \lambda}}{\omega}
=\frac{\Gamma}{L-2\Gamma} \sqrt{\frac{(x+1)(2N-x+1)}{x(2N-x)}} \, .
\end{equation}
Keeping only the largest term in the normalization condition
\mbox{$<\psi_{GGA}|\psi_{GGA}>=1$} one finds
\begin{eqnarray}
(const)^{-2}={\eta}^{2\Gamma}{\left| A(\gamma,\gamma,\gamma,\cdots,\gamma;
\omega,\omega,\omega,\cdots,\omega; 
\lambda,\lambda,\lambda,\cdots,\lambda)\right|}^2 \, .
\end{eqnarray}

The optimal value of $\Gamma$ (or $\eta$) is determined variationally
by minimizing the total energy. The energy
consists of the kinetic and the correlation part.
The kinetic energy is calculated by evaluating the
hopping term in Eq.(1). This requires the calculation of
correlation functions
\begin{equation}
\rho(i,\alpha\ ;j,\alpha)
=<\psi_{GGA}|c_{i,\alpha}^{\dagger}c_{j,\alpha}|\psi_{GGA}> \, .
\end{equation}
Because each site can be occupied by 
$x+1$, $x$, or $x-1$ electrons only, the case $i=j$
is easily evaluated to give
\begin{equation}
\rho(i,\alpha\ ;i,\alpha)=
\frac{\gamma C_{2N-1}^{x} +\omega C_{2N-1}^{x-1} +\lambda C_{2N-1}^{x-2}}{L}
=\frac{x}{2N} \, .
\end{equation}
Where, in the last step Eq.(5) is used.
For the case $i\neq j$
there are three distinct possibilities:
(a) None of two sites are doubly-occupied; (b) One of two is doubly-occupied;
(c) Both are doubly-occupied.
Summing all cases together leads to
\begin{eqnarray}
\rho(i,\alpha\ ;j\neq i,\alpha) &=&
\frac{\omega \lambda C_{2N-1}^{x-1} C_{2N-1}^{x-1} +
2\eta \omega^2 C_{2N-1}^{x-1} C_{2N-1}^{x} +
\omega^3 \lambda^{-1} C_{2N-1}^{x} C_{2N-1}^{x}}{L(L-m)} \nonumber \\
&=&\frac{\Gamma(L-2\Gamma)x(2N-x+1)}{2NL^2(2N-x)}
{\left(1+\eta \frac{\omega C_{2N-1}^{x}}{\lambda C_{2N-1}{x-1}}
\right)}^2 \\
&=&\frac{\Gamma(L-2\Gamma)}{2NL^2(2N-x)}
{\left(\sqrt{x(2N-x+1)}+\sqrt{(x+1)(2N-x)}\right)}^2 \, .\nonumber
\end{eqnarray}
In deriving these results Eq.(5) and (7) are used.
The kinetic energy per particle can be written as
\begin{equation}
K.E. = Q(x,\Gamma,N)\bar{\epsilon}(x) \, ,
\end{equation}
where
\begin{equation}
\bar{\epsilon}(x) = \frac{2}{xL} \sum_{\bf k < k_F} \epsilon({\bf k}) \leq 0
\end{equation}
is the average kinetic energy per particle
in the absence of the correlation. (The bare orbital energy has been chosen
to be zero.)
The quotient $Q(x,\Gamma,N)$,\cite{gutz} which represents
the reduction in kinetic energy of the correlated system
in comparison with that of the non-correlated system, is given by
\begin{eqnarray}
Q(x,\Gamma,N)&=&\frac{\rho(i,\alpha\ ;j\neq i,\alpha)}
{\rho(i,\alpha\ ;i,\alpha)} \nonumber \\
&=&\frac{\Gamma(L-2\Gamma)}{xL^2(2N-x)}
{\left(\sqrt{x(2N-x+1)}+\sqrt{(x+1)(2N-x)}\right)}^2 \, .
\end{eqnarray}
The correlation energy is determined by the expectation
\begin{eqnarray}
\sum_{i,\alpha\neq \beta} <\psi_{GGA}|n_{i,\alpha}^{\dagger}n_{i,\beta}|\psi_{GGA}> &=&
\Gamma C_{x+1}^{2} +\Omega C_{x}^{2} +\Lambda C_{x-1}^{2}
\nonumber \\
&=& L \frac{x(x-1)}{2}+\Gamma \,.
\end{eqnarray}
The same expectation when evaluated in the paramagnetic insulating state,
where there are $x$ electrons localized at each site, gives a value
$ L \frac{x(x-1)}{2}$. The difference, $\Gamma$,
is the increase in the correlation energy in the metallic state.
As the kinetic energy in the paramagnetic insulating state is zero,
the increase in total energy per particle in the metallic state is,
\begin{equation}
E(N,x) = Q(x,\Gamma,N)\bar{\epsilon}(x)+\frac{\Gamma}{xL}U \, ,
\end{equation}
Minimizing this energy with respect to $\Gamma$, one
obtains the fraction of doubly-occupied sites,
\begin{equation}
\frac{\Gamma}{L}=\frac{1}{4}\left(1-\frac{U}{U_c}\right) \; ,
\end{equation}
and the total energy per particle
\begin{equation}
E(x,N)=-\frac{U_c(x,N)}{8x} (1-\frac{U}{U_c(x,N)})^2 \; .
\end{equation} 
Where, the critical $U$ is given by
\begin{equation}
U_c(N,x) =\frac{\left(\sqrt{x(2N-x+1)}+
\sqrt{(x+1)(2N-x)}\right)^2}{2N-x}|\bar{\epsilon}(x)| \; .
\end{equation}
As $U$ increases toward $U_c$, the number of doubly-occupied
sites approaches zero. For $U>U_c$ the paramagnetic insulating state
has a lower energy than the metallic state.
Thus the GGA predicts a first order Mott-Hubbard metal-insulator
transition at $U_c$.

Eq.(19) is our main result. It shows that {\it $U_c$
depends sensitively on both the number of electrons per lattice
site and the band degeneracy.} 
In Fig.1 we show several examples of this dependence
with the band energy calculated from the
nearest neighbor tight-binding Hamiltonian on the simple cubic lattice.
A prominent feature is that $U_c$ is the largest at the half filling,
$U_c(N,N)=4(N+1)\bar{\epsilon}(N)$.
This implies that if the system is insulating
at the half filling, then for all other fillings with integer
number of electron per site the system is also insulating.
This result has been used successfully to explain the unusual
metallic/insulating properties of fulleride materials $A_xC_{60}$.\cite{dhub1}

The critical correlation also depends sensitively on the lattice structure
through the band energy $\bar{\epsilon}(x)$.
In Fig.2 we show examples of the phase diagram
with $\bar{\epsilon}(x)$ calculated from the 2-dimensional square,
the simple-cubic, and the body-center-cubic lattice.
As the number of nearest neighbor increases $U_c$
decreases with respect to the band width.

Our result, Eq.(19), reproduces all  
previously known analytical Gutzwiller solutions.
The three cases that we are aware of are:
(1) $N=1$, in this case the only commensurate filling is $x=1$,
Eq.(19) leads to the well known
Brinkman-Rice criteria  $U_c=8|\bar{\epsilon}|$.\cite{gutz}
(2) $N=2$ and $x=1$, this is the case studied by Chao and Gutzwiller\cite{chao} 
in early 1970s. The critical correlation they obtained,
$U_c=\frac{10+4\sqrt{6}}{3}|\bar{\epsilon}|$, is identical to that given
by Eq.(19).
(3) $N\rightarrow \infty$ with a finite $x$, in this case
the fermion problem is equivalent to the boson problem.
The boson Hubbard model has been widely
studied in recent years\cite{fisher,rokhsar,kruth}. 
Our result, $U_c=(\sqrt{x}+\sqrt{x+1})^2 | \bar{\epsilon}(x) |$,
is the same as that obtained for the boson Hubbard model
using the Gutzwiller-Approximation.\cite{rokhsar} It is
also close to the value given by quantum Monte Carlo calculations.\cite{kruth}
This suggest that in the limit of high band degeneracy our
Generalized-Gutzwiller-Approximation is quite accurate.
It would be very interesting to do Monte Carlo calculations
explicitly for the degenerate model and compare with our results.

The physical properties
of the system can undergoes dramatic changes
when the metal-insulator transition is approached.
These include the magnetic
susceptibility and the transport effective mass.
As Brinkman-Rice first pointed out,\cite{gutz} the effective mass $m^*$
is critically enhanced near the metal-insulator
transition. The enhancement is proportional to $1/Q$. Using 
Eq.(14) and (17) one finds
\begin{equation}
\frac{m^*}{m_b}=\frac{1}{Q}=\frac{8|\epsilon(x)|}
{xU_c(x,N)(1-(U/U_c)^2)} \, ,
\end{equation}
where $m_b$ is the effective band mass.
One observes that the effective mass diverges quadratically
as the metal-insulator transition is approached.
However, the enhancement is less dramatic for large $N$ or large $x$.
Brinkman-Rice also
showed that the magnetic susceptibility also diverges 
as $1/(1-(U/U_c)^2)$. We suspect that similar result
holds for the degenerate Hubbard model. However, this remains
to be proven as our calculation is valid only for
non-magnetic states.

In conclusion we present results of analytical studies on
degenerate Hubbard models using the Generalized-Gutzwiller-Approximation.
It is shown that for any filling with integer electron per lattice
site there
exists a critical correlation energy above
which the system is a Mott-Hubbard insulator.
The general expression we found for $U_c$ depends
sensitively on the band degeneracy, 
the number of electron per site and the lattice structure.
It also reproduces all previous know Gutzwiller solutions
as special limits.

\acknowledgments
This work is supported by the Petroleum Research Fund and
the U.S. Department of Energy.

\vbox{
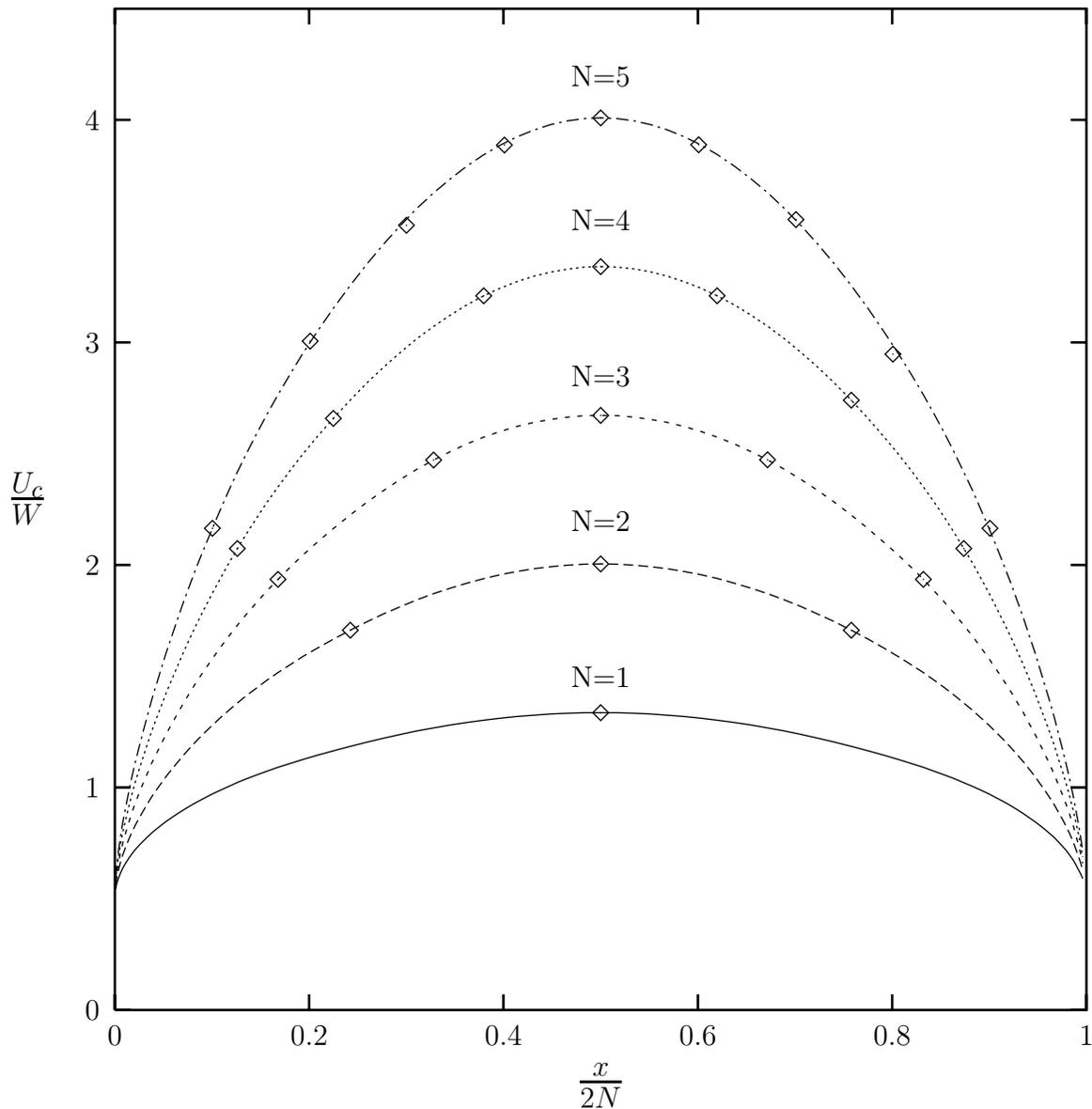
\begin{figure}
\begin{center}
\setlength{\unitlength}{0.1bp}
\begin{picture}(4679,4320)(0,0)
\put(2548,1590){\makebox(0,0){N=1}}
\put(2548,2215){\makebox(0,0){N=2}}
\put(2548,2796){\makebox(0,0){N=3}}
\put(2548,3421){\makebox(0,0){N=4}}
\put(2548,4001){\makebox(0,0){N=5}}
\put(2548,-50){\makebox(0,0){\mbox{\Large$\frac{x}{2N}$}}}
\put(250,2260){%
\makebox(0,0)[b]{\shortstack{\mbox{\Large$\frac{U_c}{W}$}}}%
}
\put(4496,151){\makebox(0,0){1}}
\put(3717,151){\makebox(0,0){0.8}}
\put(2938,151){\makebox(0,0){0.6}}
\put(2158,151){\makebox(0,0){0.4}}
\put(1379,151){\makebox(0,0){0.2}}
\put(600,151){\makebox(0,0){0}}
\put(540,3823){\makebox(0,0)[r]{4}}
\put(540,2930){\makebox(0,0)[r]{3}}
\put(540,2037){\makebox(0,0)[r]{2}}
\put(540,1144){\makebox(0,0)[r]{1}}
\put(540,251){\makebox(0,0)[r]{0}}
\end{picture}
\end{center}
\vskip0.1in
\caption{
Phase diagrams of metal-insulator transitions
in degenerate Hubbard models as predicted by Eq.(19). Critical correlation
$U_c$ (scaled by the band width) is plotted
against the band filling $x$ (scaled by $2N$). The band energy is calculated
from the n.n. tight-binding model on the simple-cubic lattice.
Shown are results for $N=1,2,3,4,5$. The $U_c$ is the largest
at the half-filling $x=N$.
In both Fig.1 and Fig.2 only points, corresponding to
integer number of electrons per site, are meaningful.
Lines are draw using Eq.(19).
}
\end{figure}
}

\vbox{
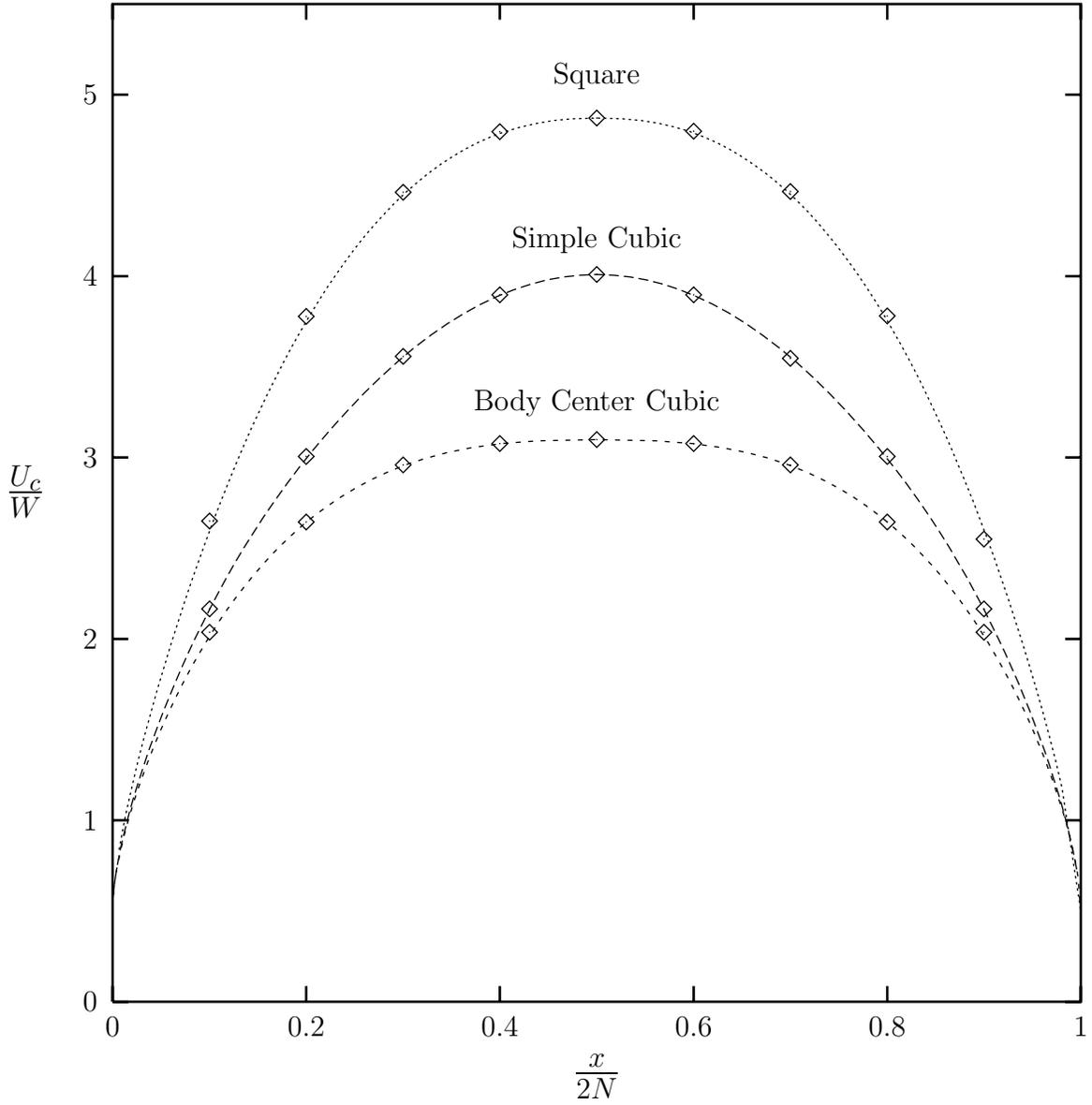
\begin{figure}
\begin{center}
\setlength{\unitlength}{0.1bp}
\begin{picture}(4679,4320)(0,0)
\put(2548,2662){\makebox(0,0){Body Center Cubic}}
\put(2548,3319){\makebox(0,0){Simple Cubic}}
\put(2548,3977){\makebox(0,0){Square}}
\put(2548,-50){\makebox(0,0){\mbox{\Large$\frac{x}{2N}$}}}
\put(250,2260){%
\makebox(0,0)[b]{\shortstack{\mbox{\Large$\frac{U_c}{W}$}}}%
}
\put(4496,151){\makebox(0,0){1}}
\put(3717,151){\makebox(0,0){0.8}}
\put(2938,151){\makebox(0,0){0.6}}
\put(2158,151){\makebox(0,0){0.4}}
\put(1379,151){\makebox(0,0){0.2}}
\put(600,151){\makebox(0,0){0}}
\put(540,3904){\makebox(0,0)[r]{5}}
\put(540,3173){\makebox(0,0)[r]{4}}
\put(540,2443){\makebox(0,0)[r]{3}}
\put(540,1712){\makebox(0,0)[r]{2}}
\put(540,982){\makebox(0,0)[r]{1}}
\put(540,251){\makebox(0,0)[r]{0}}
\end{picture}

\end{center}
\vskip0.1in
\caption{
Phase diagrams of metal-insulator transitions in
degenerate Hubbard models for different lattice structures.
Shown are three examples with $N=5$ and the band structure
calculated for the 2-D square, 3-D simple-cubic, and 3-D body-center-cubic
lattices with the nearest neighbor
tight-binding hamiltonian.
Note that given the same band filling and the degeneracy,
the critical correlation can changes substantially from one type
of lattice structure to another.
}
\end{figure}
}
\end{document}